\documentclass[12pt,a4paper]{article}

\usepackage[cp1251]{inputenc}
\usepackage[english]{babel}

\usepackage{indentfirst}
\usepackage{amsmath}
\usepackage{amssymb}
\usepackage[pdftex]{graphicx,color}
\usepackage{slashed}

\usepackage{geometry}
\begin{document}

\begin{flushright}
SI-HEP-2017-07 

QFET-2017-06 \\[0.2cm]
\end{flushright}


\begin{center}
\textbf{\Large  Higher-Twist Effects in Light-Cone Sum Rule for 
the~$B\to\pi$ Form Factor}

\vspace*{1.6cm}

{\large Aleksey~V.~Rusov}
 
\vspace*{0.4cm}

\textsl{%
Theoretische Physik 1,
Naturwissenschaftlich-Technische Fakult\"at,\\
Universit\"at Siegen, D-57068 Siegen, Germany}
\vspace*{0.2cm}

\textsl{%
Department of Theoretical Physics,
P.G. Demidov Yaroslavl State University, \\ 150000, Yaroslavl, Russia}
\vspace*{0.8cm}

\textbf{Abstract}\\[10pt]
\parbox[t]{0.9\textwidth}{
We calculate the higher-twist corrections to the 
QCD light-cone sum rule for the $B \to \pi$ transition form factor. 
The light-cone expansion of the massive quark propagator in the external 
gluonic field is extended to include new terms containing the derivatives of 
gluon-field strength. The resulting analytical expressions for the twist-5 and twist-6 
contributions to the correlation function are obtained in a factorized 
approximation, expressed via the product of the lower-twist pion distribution amplitudes  
and the quark-condensate density. 
The numerical analysis reveals that new higher-twist effects for the $B \to \pi$ 
form factor are strongly suppressed.
This result justifies the conventional truncation of the operator product expansion 
in the light-cone sum rules up to twist-4 terms.}

\end{center}

\vspace*{3cm}

\newpage

\section{Introduction}

Accurate calculation of the $B \to \pi$ transition form factors in QCD plays 
an important role, since, for instance, the vector form factor is used for the determination of 
the Cabibbo-Kobayashi-Maskawa (CKM) matrix element $V_{ub}$ from the experimental data on 
the exclusive $B \to \pi \ell \nu_\ell$ decays.
The $B \to \pi$ transition form factors are nonperturbative quantities accounting for the
complicated quark-gluon dynamics inside the meson states and
can be calculated using different QCD-based approaches. Among them,
the method of light cone sum rules (LCSR) \cite{Balitsky:1986st, Chernyak:1990ag}
is applicable at large hadronic recoil \cite{Belyaev:1993wp, Belyaev:1994zk}. 
The main advantage of this method is the possibility to 
perform calculation in full QCD, with finite $b$-quark mass.
The starting object of the calculation is a properly designed
correlation function of the quark currents for which the operator product expansion (OPE)
near the light-cone is applicable. 
Within OPE, the correlation function is decomposed into a series of the
hard-scattering kernels convoluted with the pion light-cone distribution amplitudes (DA's)
of the growing twist. The result of the OPE for correlation function 
is related to the $B \to \pi$ form factor employing the hadronic dispersion relation
and quark-hadron duality.

At present time the accuracy of the LCSR calculation of heavy-to-light transition form factors 
is limited by the contributions of the operators up to twist 4.
The results for the relevant partial contributions of the twist-2, -3 and -4 terms to the
LCSR as well as radiative gluon corrections to the corresponding 
hard-scattering kernels of the twist-2 and twist-3 terms can be found in
\cite{Belyaev:1993wp, Belyaev:1994zk, Duplancic:2008ix, Khodjamirian:1997ub, Bagan:1997bp, Ball:2001fp}. 
Moreover, a $\beta_0$ estimation for the twist-2 $O (\alpha_s^2)$ 
contributions can be found in \cite{Bharucha:2012wy}.
It is important to note that the contributions of even- and odd-twist terms in the OPE form 
two separate hierarchies with respect to the lowest twist-2 and twist-3 terms, respectively.
Note also that the twist-3 term, despite power suppression,
contains a chirally enhanced parameter $\mu_\pi = m_\pi^2 / (m_u + m_d)$, 
which renders the twist-3 contribution to the same order of magnitude as the twist-2 one.
The contribution of twist-4 term was found to be
significantly suppressed in comparison with the corresponding twist-2 one
\cite{Duplancic:2008ix}.
Such a comparison in the odd-twist hierarchy is still not possible
due to missing estimate of twist-5 effects. 
Moreover, an estimate of the twist-6 term contribution to LCSR
will allow us to confirm the expected power suppression of the  
higher twists in the even-twist hierarchy.
The main purpose of this work is to evaluate the twist-5 and twist-6 contributions 
to the LCSR for the $B \to \pi$ form factors.

The calculation of the higher twist effects in the OPE  near the light-cone 
is interesting for several reasons. As mentioned in Ref.~\cite{Braun:1999uj}, 
the twist-3 and twist-4 operators cannot be factorized as a product of 
the gauge invariant operators of lower twist.
There are several operators of twist 5 and twist 6 which can be factorized
as a product of the gauge-invariant operators of lower twist.
Sandwiched between the vacuum and one-pion state, such operators generally produce
two types of contributions: factorizable ones in terms of a lower-twist 
two-particle distribution amplitude times quark condensate
and nonfactorizable ones, which give rise to genuine twist-5 and twist-6 multiparton
pion distribution amplitudes.
As argued in \cite{Braun:1999uj}, in the context of conformal symmetry 
the contributions of higher Fock states are strongly suppressed 
and their contributions to the sum rules are probably negligible.
Factorizable contributions, on the other hand, can be comparatively large.
Hence their calculation practically solves the problem of investigating
the OPE beyond the twist-4 level. 

In \cite{Braun:1999uj} and \cite{Agaev:2010aq}
the factorizable twist-6 contributions in LCSR's for the pion electromagnetic
and $\pi \gamma^* \gamma$ form factors, respectively, were computed.
In fact, in these sum rules the twist-6 contributions are the only ones 
which arise in the presence of virtual massless ($u$- or $d$)-quark in the correlation 
function, hence, only the even twists are relevant there. 
Here we extend the analogous calculation  to the correlation function with 
a massive virtual quark. 
In this case both factorizable twist-5 and twist-6 terms contribute to LCSR. 
In order to obtain these contributions one needs the massive quark propagator expanded near 
the light-cone up to the terms including the derivatives of the gluon field strength.   
The analytical expression for this propagator as well as the factorizable twist-5 and twist-6
contributions to LCSR represent new results obtained here.

The paper is organised as follows.
Sec.~2 is devoted to the derivation of the new terms in the expansion of the massive 
quark propagating in the external gluonic field near the light-cone.
In Sec.~3 the detailed calculation of the diagrams corresponding to 
the factorizable twist-5 and twist-6 contributions to the LCSR for the 
vector $B \to \pi$ form factor is presented.
Sec.~4 contains the relevant numerical estimates and Sec.~5 the concluding discussion.
Some useful formulae are collected in the appendix.

\section{Light-cone expansion of the massive quark propagator in the external gluon field}
For our purpose we need the light-cone (LC) expansion of the quark propagator in the external 
gluon field. The corresponding expression including the terms with the covariant derivatives 
of gluon field strength is known only in the case of massless quark
and was derived for instance in \cite{Balitsky:1987bk} (see also \cite{Braun:1999uj}).
For a {\it massive} quark propagator the corresponding result is known only at leading order 
of the LC-expansion in the gluon field. 
To estimate the higher twist effects in the $B \to \pi$ form factors we need 
also to include the higher order terms in LC-expansion which are proportional to
the covariant derivatives of the gluon-field strength. 
This task is technically more involved due to a presence of the quark mass~$m$. 

In order to get the LC-expansion of the massive quark propagator up to the needed accuracy 
we start from the definition of the quark propagator:
\begin{equation}
S (x,x^\prime) = - i \langle 0 | T\{\psi (x), \bar \psi (x^\prime)\} | 0 \rangle,
\end{equation} 
where $\psi (x)$ denotes the massive quark field operator.
Hereafter we choose $x^\prime = 0$ for simplicity.
The propagator satisfies the usual Green-function equation
\begin{equation}
\left( i \gamma^\mu \partial_\mu + g_s \gamma^\mu A_\mu (x)  - m \right)
S(x,0) = \delta^{(4)} (x),
\label{eq:eq-for-prop}
\end{equation}
where $A_\mu = A_\mu^a \lambda^a/2$ is the four-potential of the gluon field, and
$\lambda^a$ are the Gell-Mann matrices $(a = 1, \ldots 8)$.
The solution of (\ref{eq:eq-for-prop}) can be presented in the form of perturbative series  
in the power of the strong coupling $g_s$:
\begin{equation}
i S(x,0) = i S^{(0)} (x) + i S^{(1)} (x) + \ldots
\label{eq:prop-series}
\end{equation}
where 
\begin{equation}
i S^{(1)} (x) = g_s \int d^4 y \, i S^{(0)} (x-y) \, i \slashed A (y) \, i S^{(0)} (y), 
\end{equation}
and $S^{(0)}$ denotes the free quark propagator.
The four-potential of the gluon field is taken in the Fock-Schwinger (of fixed point) gauge, 
so that $(x_\mu - x^\prime_\mu) A^\mu (x) = 0$ and $x^\prime = 0$. 
For further calculation it is convenient to use the free quark propagator $S^{(0)} (x-y)$
in the form of so-called $\alpha$-representation
\begin{equation}
S^{(0)} (x-y) = - i \int\limits_0^\infty \frac{d \alpha}{16 \pi^2 \alpha^2} 
\left(m + i \frac{\slashed x - \slashed y}{2 \alpha} \right) 
e^{-m^2 \alpha + \frac{(x-y)^2}{4 \alpha}}, 
\label{eq:mas-prop-alpha-repres}
\end{equation}
which allows us to rewrite the first order correction $S^{(1)} (x,0)$ 
to the propagator as follows:
\begin{eqnarray}
\label{eq:prop-first-step}
S^{(1)} (x,0) & =  & \frac{g_s}{(16 \pi^2)^2} \int\limits_0^\infty \frac{d\alpha}{\alpha^2} \int\limits_0^\infty \frac{d \beta}{\beta^2} \int d^4 y \\
& \times & \left( m + i \frac{\slashed x - \slashed y}{2 \alpha} \right)
\slashed A (y) \left( m + i \frac{\slashed y}{2 \beta} \right)
e^{-m^2 (\alpha + \beta)} e^{\frac{(x-y)^2}{4\alpha}+ \frac{y^2}{4\beta}}.
\nonumber  
\end{eqnarray}
Transforming the integration variable $\beta$ as:
\begin{equation}
\beta = \frac{\alpha u}{1-u}, \quad 0 < \beta < \infty \quad \mbox{ so that } \quad 0 < u < 1,
\label{eq:beta-to-u}
\end{equation} 
we introduce a new variable:
\begin{equation}
y^\prime = y - u x. 
\label{eq:y-to-primey}
\end{equation}
Taking into account the replacements (\ref{eq:beta-to-u}) and (\ref{eq:y-to-primey})
one can represent the expression (\ref{eq:prop-first-step}) in the form 
(hereafter we redefine $y^\prime \to y$):
\begin{eqnarray}
\lefteqn{S^{(1)} (x,0) = \frac{g_s}{(16 \pi^2)^2} \int\limits_0^1 \frac{d u}{u^2} 
\int\limits_0^\infty \frac{d \alpha}{\alpha^3} \int d^4 y \, e^{-m^2 \alpha /\bar u} 
e^{[y^2 + x^2 u \bar u]/(4 \alpha u)} }
\label{eq:prop-second-step} \\
 & \times & \Bigg\{m^2 \slashed A (y + u x) + \frac{i m}{2 \alpha u} 
\Big[2 u \bar u (x \cdot A(y + u x)) 
- 2 u (y \cdot A(y + u x)) + \slashed A (y + u x) \slashed y \Big] -
\nonumber \\
& - & \frac{\bar u}{4 \alpha^2 u} \Big[u \bar u \slashed x \slashed A (y + u x) \slashed x 
- u \slashed y \slashed A (y + u x) \slashed x 
+ \bar u \slashed x \slashed A (y + u x) \slashed y 
- \slashed y \slashed A (y + u x) \slashed y \Big] \Bigg\},
\nonumber
\end{eqnarray}
where $\bar u = 1 -u$, $(x (y) \cdot A) \equiv x_\mu (y_\mu) A^\mu$.
After that we expand the field $A_\alpha (y + u x) $ in the powers of 
the deviation $y_\mu$ from the point $ u x $ near the light cone 
$(x^2 \simeq 0)$:
\begin{eqnarray}
A_\alpha (y + u x) & = & A_\alpha (u x) + \partial_\mu A_\alpha (u x) y^\mu + \frac{1}{2} \partial_\mu \partial_\nu A_\alpha (u x) y^\mu y^\nu \nonumber \\
& + & \frac{1}{6} \partial_\mu \partial_\nu \partial_\rho A_\alpha (u x) y^\mu y^\nu y^\rho 
+ \ldots,
\label{eq:A-expan}
\end{eqnarray}
with the following shorthand notation:
$
\partial_\mu A_\alpha (u x) \equiv 
\left. \frac{\partial A_\alpha (z)}{\partial z^\mu}\right|_{z = u x}.
$
Substituting the expansion (\ref{eq:A-expan}) in (\ref{eq:prop-second-step})
allows to calculate $S^{(1)} (x,0)$ order by order.
Performing the Wick's rotation $y_0 \to - i y_4$, one reduces the integrals over $d^4 y$ 
to the standard Gaussian integrals. 
After integration over $d^4 y$, one calculates the integrals over $\alpha$ 
introducing the modified Bessel function of the second kind $K_n (z)$:
\begin{equation}
\label{eq:BesselK-prop}
\int\limits_0^\infty \frac{d \alpha}{\alpha^n} \, {\rm exp} 
\left[- \frac{m^2 \alpha}{1-u} + \frac{x^2 (1-u)}{4 \alpha} \right] 
= 2 \left( \frac{2 m}{\sqrt{-x^2} (1-u)} \right)^{\! \! \frac{n-1}{2}} 
\! \! \! K_{n-1} (m \sqrt{-x^2}), \quad x^2 < 0.
\end{equation}
Then we perform some transformations in order to relate 
the derivatives of $A_\rho (xu)$ with $G^{\mu \nu} (u x)$ and its derivatives. 
The first term in the expansion (\ref{eq:A-expan}) yields the scalar product
$(x \cdot A)$ which vanishes in the Fock-Schwinger gauge.
Since $S^{(1)} (x,0)$ has ${\cal O} (g_s)$ accuracy, the partial derivatives
$\partial_\mu$ can be replaced by the covariant ones $D_\mu$.
Taking into account the definition of the gluon-field strength tensor 
$G_{\mu \nu} = G_{\mu \nu}^{a} \lambda^a /2 = D_\mu A_\nu - D_\nu A_\mu$, 
one relates then the covariant derivatives of $A_\mu$ with $G_{\mu \nu}$ and its derivatives.
We found that the terms proportional to $D^\mu A_\mu$ vanish after integration 
by parts in variable $u$, allowing one to present the final result for the propagator 
in terms of gluon-field strength only.

After lengthy but straightforward calculation we arrive at the following expression for 
the massive quark propagator expanded near the light-cone, including terms up to 
the second derivative of the gluon field strength:
\footnote{This form of the propagator has been derived in the space-like region of $x^2$. Performing similar calculations for positive $x^2$ one can demonstrate 
that the propagator is expressed via the Hankel functions of the second kind $H_n^{(2)}(z)$.
Nevertheless, the representation (\ref{eq:prop-final-exp-x}) can be also used 
for positive $x^2$ having in mind the following relation between these special functions 
$$
K_n (i z) = \frac{\pi}{2} (- i)^{n+1} H_n^{(2)} (z), \quad z > 0,
$$
allowing to continue Bessel functions $K_n (m \sqrt{-x^2})$ 
to the positive $x^2$-domain.
} 
\begin{eqnarray}
S(x,0) & = & 
- \frac{i m^2}{4 \pi^2} \left[ \frac{K_1 (m \sqrt{-x^2})}{\sqrt{-x^2}} +
i \frac{\slashed x}{-x^2} K_2(m \sqrt{-x^2}) \right]   
\label{eq:prop-final-exp-x} \\
& - & 
\frac{i g_s}{16 \pi^2} \int \limits_0^1 d u \Biggl[
m K_0 (m \sqrt{-x^2}) (G (u x) \cdot \sigma) 
\nonumber \\
& + & 
\frac{i m}{\sqrt{-x^2}} K_1 (m \sqrt{-x^2})
\left[ \bar u \slashed x (G (u x) \cdot \sigma) + u (G (u x) \cdot \sigma) \slashed x \right] 
\nonumber \\
& - & 
2 u \bar u \left(i m K_0 (m \sqrt{-x^2}) - \frac{m \slashed x}{\sqrt{-x^2}} 
K_1 (m \sqrt{-x^2}) \right) x_\mu D_\nu G^{\nu \mu} (u x) 
\nonumber \\
& + & 
K_0 (m \sqrt{-x^2}) \left(2 u \bar u - 1 \right) \gamma_\mu D_\nu G^{\nu \mu} (u x)  
\nonumber \\
& - & 
u \bar u (1 -2 u) K_0 (m \sqrt{-x^2}) x_\mu \slashed D D_\nu G^{\nu \mu} (u x)  
\nonumber \\
& - & 
i u \bar u K_0 (m \sqrt{-x^2}) \epsilon_{\sigma \mu \nu \rho} x^\sigma 
\gamma^\mu \gamma_5 D^\nu D_\alpha G^{\alpha \rho} (u x)  
\nonumber \\
& + & 
u \bar u \sqrt{-x^2} K_1 (m \sqrt{-x^2})
 \sigma_{\rho}^{\phantom{\rho} \nu} D_\nu D_\mu G^{\mu \rho} (u x) \Biggr] + \ldots
\nonumber
\end{eqnarray}
where $(G \cdot \sigma) \equiv G_{\mu \nu} \sigma^{\mu \nu}, \,
\sigma^{\mu\nu} = (i/2) [\gamma^\mu, \gamma^\nu]$, 
and dots denote the higher powers of the light-cone expansion of $G_{\mu \nu}$ 
and corrections with two and more gluons, which are beyond the approximation we need.
Taking into account the asymptotics of the Bessel functions:
\begin{eqnarray}
\left. K_0 (m \sqrt{-x^2}) \right|_{m\to 0} & \sim & - \gamma_E - \ln \left(\frac{m}{2}\right)
-\frac{1}{2} \ln \left( -x^2 \right),
\nonumber \\
\left. \sqrt{-x^2} K_1 (m \sqrt{- x^2}) \right|_{m\to 0} & \sim & \frac{1}{m},
\end{eqnarray}
one reproduces the corresponding result in the case of the massless quark 
given in \cite{Balitsky:1987bk, Braun:1999uj}.

We also found that the resulting expression (\ref{eq:prop-final-exp-x}) can be rewritten in an equivalent Fourier-transformed form:
\begin{eqnarray}
S(x, 0) & = & \int \frac{d^4 p}{(2 \pi)^4} e^{-i p x} \Biggl\{ \frac{\slashed p + m}{p^2 - m^2} 
- \frac{g_s}{\left(p^2 - m^2 \right)^3} \int\limits_0^1 d u \biggl[ 
\frac{1}{2} m (p^2 -m^2) (G (u x) \cdot \sigma) + 
\nonumber \\
& + & \frac{1}{2} (p^2 - m^2)
\left( \bar u \slashed p (G (u x) \cdot \sigma) + u (G (u x) \cdot \sigma) \slashed p \right)
- 4 u \bar u (\slashed p + m) p_\mu D_\nu G^{\nu \mu}(u x) - 
\nonumber \\
& - & \frac{1}{2} (p^2 - m^2) \gamma_\mu D_\nu G^{\nu \mu} (u x)  - 2 i u \bar u (1 - 2u)
p_\mu \slashed D D_\nu G^{\nu \mu} (u x) + 
\nonumber \\
& + &
2 u \bar u \epsilon_{\sigma \mu \nu \rho} p^{\sigma} \gamma^\mu \gamma_5 
D^\nu D_{\alpha} G^{\alpha \rho} (u x) - 
2 m u \bar u \sigma_{\rho}^{\phantom{\rho} \nu} D_\nu D_\mu G^{\mu \rho} (u x)
+\ldots \biggr] \Biggr\}.
\label{eq:prop-final-exp-p}
\end{eqnarray}
The first terms of this expression are in full agreement with the LC-expansion 
of the massive quark propagator given in \cite{Belyaev:1994zk}, 
and the terms with the covariant derivative of the gluon field strength represent
a new result of this paper.

\section{Factorizable twist-5 and twist-6 contributions to the $B~\to~\pi$ form factor}

The starting object for a calculation of the $B \to \pi$ form factors in the framework
of the LCSR approach is the following correlation function of the 
$B$-meson interpolating and the $b \to u$ weak transition currents:
\begin{eqnarray}
F_\mu (p, q) & = & i \int \! d^4 x \, e^{i q x} 
\langle \pi (p) | T \{\bar u (x) \gamma_\mu b(x), m_b \bar b(0) i \gamma_5 d (0)\} 
|0 \rangle \nonumber \\ 
& = &  F (q^2, (p+q)^2) p_\mu + \tilde{F} (q^2, (p+q)^2) q_\mu,
\label{eq:corr-func-def}
\end{eqnarray}
where $p$ is the four-momentum of the pion, $q$ is the outgoing four-momentum,
and $m_b$ is the $b$-quark mass.
For definiteness, we consider the $\bar B_d^0 \to \pi^+$ flavour configuration. 
The Lorentz-invariant amplitudes $F (q^2, (p+q)^2)$ 
and $ \tilde F (q^2, (p+q)^2)$ are used for the calculation of the
vector and scalar form factors.
In this paper we focus on an estimate of the higher twist effects for the vector 
form factor, hence, we need to consider only the amplitude $F (q^2, (p+q)^2)$.
In the framework of LCSR approach one considers the correlation function 
(\ref{eq:corr-func-def}) in the kinematic domain $q^2 \ll m_b^2$ and 
$(p+q)^2 \ll m_b^2$, far from the $b$-flavour threshold.
In this domain the separations near the light-cone dominate 
and one can expand the integrand in (\ref{eq:corr-func-def}) near $x^2 = 0$
(see e.g. \cite{Duplancic:2008ix}).
Contracting the virtual $b$-quark fields one rewrites 
(\ref{eq:corr-func-def}) in the form
\begin{equation}
\label{eq:correl-func-step-2}
F_\mu (p,q) = - m_b \int d^4 x \, e^{i q x} \langle \pi (p) |
\bar u (x) \gamma_\mu i S_b (x,0) \gamma_5 d(0) | 0 \rangle, 
\end{equation}
where $S_b (x,0)$ denotes the $b$-quark propagator expanded near the light-cone.

Currently, the accuracy of the OPE for the correlation function at leading order 
in $\alpha_s$ is limited by contributions up to twist-4 terms.
In our paper, we focus on a derivation of the factorizable 
twist-5 and twist-6 contributions. To this end,
we substitute the LC-expansion of the $b$-quark propagator calculated 
in the previous section (see eq.~(\ref{eq:prop-final-exp-x})) and take only terms
proportional to the derivative $D_\mu G^{\mu \nu}$ of the gluon-field strength. 
The latter are transformed by applying the equation of motion for 
the gluon-field strength:
\begin{equation}
\label{eq:eom}
D_\mu G^{\mu \nu} (u x) = -g_s \sum_{q} 
\left(\bar q (u x) \gamma^\nu \frac{\lambda^a}{2} q(u x)\right)\frac{\lambda^a}{2}.
\end{equation}
In the above, due to the quark content of the final state pion, only the terms 
with $u$ and $d$-quark contribute. 
Applying the equation of motion (\ref{eq:eom}) yields the matrix elements
of two quark-antiquark operators sandwiched between pion and vacuum states.
These matrix elements generate two different types of contributions.
The first ones related to the four-particle DAs are expected to be negligible
\cite{Braun:1999uj}.
On the other hand, the contributions of the second type (factorizable) could have 
larger numerical impact on LCSR for the form factor.
In this paper following the same approach as in \cite{Braun:1999uj, Agaev:2010aq}
we restrict ourselves by the factorization approximation  and present 
the matrix elements of the two quark-antiquark operators
as a product of the dimension-three quark condensate 
$\langle \bar q  q \rangle$ and the bilocal vacuum-pion matrix element containing
pion twist-2 and twist-3 light-cone distribution amplitudes (LCDA's). 
The latter matrix element can be presented in the form \cite{Duplancic:2008ix}:
\begin{eqnarray}
\langle \pi (p) | \bar u_\alpha^i (x_1) d_\beta^j (x_2) | 0 \rangle \Big|_{(x_1-x_2)^2 \to 0} & = & \frac{i \delta^{ij} f_\pi} {12} \int\limits_0^1 \! dv \, e^{i v (p x_1) + i \bar v( px_2)} 
\biggl( [\slashed p \gamma_5]_{\beta \alpha} \, \varphi (v) 
\label{eq:twist-2-3-matr-elem} \\
& - & [\gamma_5]_{\beta \alpha} \, \mu_\pi \varphi_p (v) 
+ \frac{1}{6} [\sigma_{\mu \nu} \gamma_5]_{\beta \alpha} 
\, p^\mu (x_1 - x_2)^\nu \mu_\pi \varphi_\sigma (v) \biggr),
\nonumber
\end{eqnarray}
where the upper $i,j$ and lower $\alpha, \beta$ indices are the colour 
and bispinor indices of the quark fields, respectively, 
$\bar v = 1 - v$,  $f_\pi$ is the pion decay constant, and $\varphi (v)$ 
and $\varphi_{p, \sigma} (v)$ denote the twist-2 and twist-3 pion 
light-cone DA's, respectively.

\begin{figure}[t]\center
\includegraphics[scale=0.4]{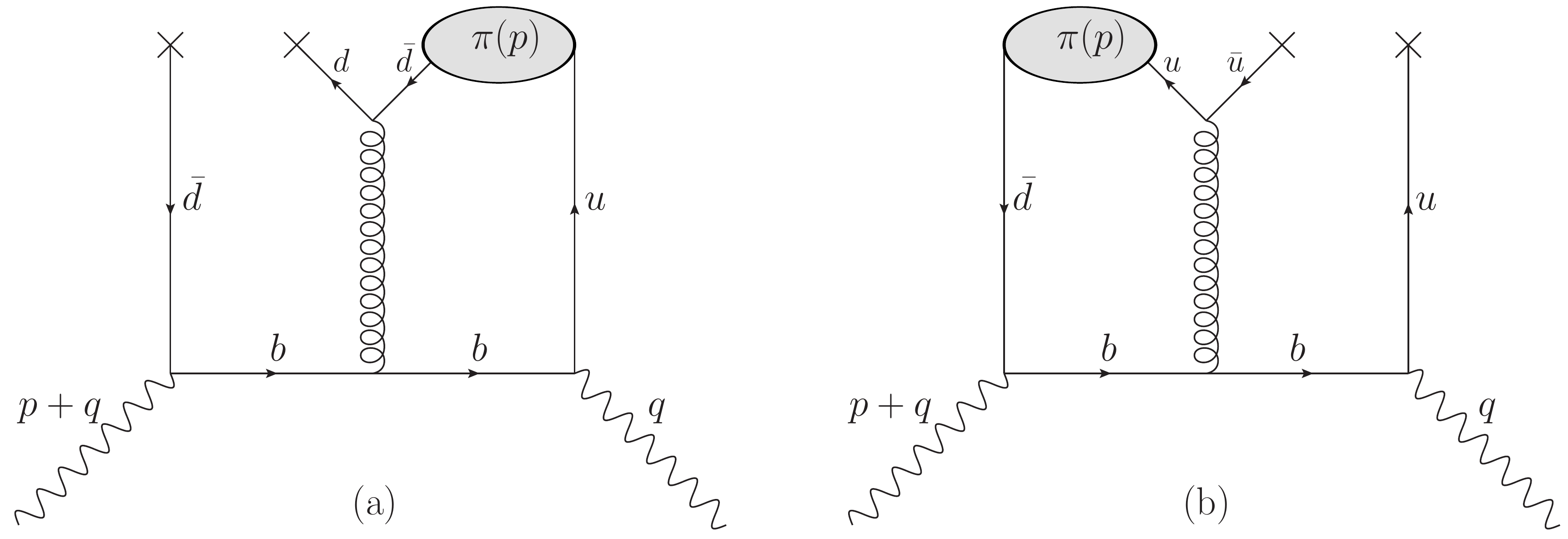}
\caption{Diagrams representing the factorizable twist-5 and twist-6 contributions 
to the correlation function (\ref{eq:corr-func-def}).}
\label{fig:diag-tw-5-6}
\end{figure}

The matrix elements of the two quark fields sandwiched between the vacuum states 
can be expressed via the quark vacuum condensate in the local limit $|x_1 - x_2| \to 0$. 
Expanding the light quark field $q (x) = u (x)$ or $d(x)$ near the point $x = 0$ 
one can demonstrate that~\cite{Colangelo:2000dp}:  
\begin{equation}
\label{eq:matr-elem-to-cond-appr}
\langle 0 | \bar q_\alpha^i (x) q_\beta^j (0) | 0 \rangle \simeq \frac{\delta^{ij} 
\delta_{\alpha\beta}}{12} \langle \bar q q \rangle, 
\end{equation}
where $\langle \bar q q \rangle$ denotes the dimension-3 light quark condensate,
and we assume isospin symmetry, therefore $\langle \bar q q \rangle \equiv
\langle \bar u u \rangle = \langle \bar d d \rangle$.
The corresponding contributions of the factorizable twist-5 and twist-6
terms to the OPE for the correlation function are described by diagrams 
shown in Fig.~\ref{fig:diag-tw-5-6}.
They are formed only by the gluon emitted from the virtual $b$-quark.
Gluons emitted from the light $\bar d$ and $u$ quarks and converted
to the quark-antiquark pair represent a genuine long-distance effect 
which is by default included in the DA's.
Such an implicit separation of long- and short-distance effects takes place also 
in the diagrams with three-particle quark-antiquark-gluon DA's of twist 3,4.

After factorization the further calculation is straightforward albeit lengthy. 
The final result of the OPE for the correlation function reads:
\begin{eqnarray}
\lefteqn{F^{\rm (OPE)}_{\rm tw5,6}(q^2,(p+q)^2) =  
\alpha_s \langle \bar q q \rangle \frac{C_F }{N_c} \pi m_b f_\pi 
\int\limits_0^1 du \int\limits_0^1 d v}
\label{eq:F-tw-5-6} \\ 
& \times & \Biggl\{\varphi (v) \Biggl[
\frac{1}
{[m_b^2 - (q + u v p)^2]^2} + 
\frac{2 (1 - 2 u \bar u)}
{[m_b^2 - (q + (u + v - u v)p)^2]^2} 
\nonumber \\
& + &
\frac{4 u \bar u \, q^2}
{[m_b^2 - (q + u v p)^2]^3} + 
\frac{4 u \bar u \, m_b^2}
{[m_b^2 - (q + (u + v - u v)p)^2]^3} 
\Biggr]
\nonumber \\
& - & 4 \mu_\pi m_b \, \varphi_p (v) \Biggl[
\frac{u^2 \bar u v }
{[m_b^2 - (q + u v p)^2]^3} + 
\frac{u \bar u (u v - u -v)}
{[m_b^2 - (q + (u + v - u v)p)^2]^3} \Biggr]
\nonumber \\
& + & 2 \mu_\pi m_b \, \varphi_\sigma (v) \Biggl[
\frac{u^2 \bar u (m_b^2 + q^2)}
{[m_b^2 - (q + u v p)^2]^4} + 
\frac{u \bar u^2 (m_b^2 + q^2)}
{[m_b^2 - (q + (u + v - u v)p)^2]^4} \Biggr]
\Biggr\},
\nonumber
\end{eqnarray}
where the contributions of factorizable twist-5 and twist-6 terms are separated.
Note, that in the above the masses of the pion and light quarks are neglected, $m_\pi = 0$ 
and $m_{u,d} = 0$, everywhere except in the parameter $\mu_\pi = m_\pi^2/(m_u + m_d)$. 

In order to estimate the corresponding correction to the vector $B \to \pi$ form factor 
one follows the standard procedure of the LCSR derivation.
First of all, one needs to perform the change of the integration variables 
in (\ref{eq:F-tw-5-6}) in order to present the OPE result for the invariant amplitude 
$F(q^2, (p+q)^2)$ as a quasi-dispersion integral in the variable $(p+q)^2$.
One obtains:
\begin{equation}
F^{\rm (OPE)}_{\rm tw5,6} (q^2, (p+q)^2) =
\alpha_s \langle \bar q q \rangle \frac{C_F }{N_c} \pi m_b f_\pi 
\int\limits_{m_b^2}^\infty d s \sum_{n = 2,3,4} \frac{g_n (q^2, s)}{(s-(p+q)^2)^n},
\label{eq:OPE-disp-form}
\end{equation}
where the details of derivation and the explicit expressions of functions $g_n (q^2,s)$ 
are given in Appendix.

To access the vector $B \to \pi$ form factor, one writes down the hadronic 
dispersion relation for the invariant amplitude $F (q^2, (p+q)^2)$ in the channel of 
the $\bar b \gamma_5 d$ current with the four-momentum squared $(p+q)^2$.
Inserting a full set of the hadronic states with quantum numbers of $B$-meson
between the currents in (\ref{eq:corr-func-def})
one isolates the ground state $B$-meson contribution in the dispersion integral.
To this end, we need to define the hadronic matrix elements:
\begin{equation}
i m _b \langle B |\bar b \gamma_5 d |0 \rangle = m_B^2 f_B,
\label{eq:ME-decay-constant}
\end{equation}
\begin{equation}
\langle \pi (p) | \bar q \gamma^\mu b | B (p+q) \rangle =  
f^+_{B \pi}  (q^2) \left [2 p^\mu + \left(1 - \frac{m_B^2-m_\pi^2}{q^2} \right)q^\mu \right] 
+ f^0_{B \pi}  (q^2) \, \frac{m_B^2-m_\pi^2}{q^2} \, q^\mu,  
\label{eq:fp-and-f0-def}
\end{equation}
where $f_B$ is the $B$-meson decay constant and $f^+_{B\pi} (q^2) $ and 
$f^0_{B\pi} (q^2)$ are the standard $B \to \pi$ vector and scalar form factors.  
One presents then the amplitude $F (q^2, (p+q)^2)$ as follows:
\begin{equation}
F (q^2, (p+q)^2) = \frac{2 f_{B\pi}^+ (q^2) m_B^2 f_B}{m_B^2 - (p+q)^2}
+ \int\limits_{s_0^B}^{\infty} d s \frac{\rho^h (q^2, s)}{s-(p+q)^2}.
\label{eq:F-Hadr-disp-rel}
\end{equation}
In the above, the contribution of the excited states and continuum of hadrons 
with the same quantum numbers as $B$-meson is presented in the form of 
the integral over the spectral density  $\rho^h (q^2,s)$.
Its contribution can be related with the OPE result by means of the quark-hadron duality
\begin{equation}
\rho^h (q^2, s) = \frac{1}{\pi} {\rm Im} F^{\rm (OPE)} (q^2, s) \, \Theta (s-s_0^B),
\label{eq:quark-hadr-dual}
\end{equation}
introducing the effective continuum threshold $s_0^B$.
The imaginary part of the invariant amplitude 
${\rm Im} F^{\rm (OPE)} (q^2, (p+q)^2)$ in the variable
$(p+q)^2$ is easily extracted from (\ref{eq:OPE-disp-form}).
In order to suppress the contribution of  the excited states 
one applies the Borel transformation, replacing the variable $(p+q)^2$ 
by the Borel parameter $M^2$.
Finally, after substraction of the continuum contribution 
the corresponding twist-5 and twist-6 corrections for the vector $B \to \pi$ form factor 
can be presented in the following compact form:
\begin{equation}
\left[f_{B\pi}^+ (q^2)\right]_{\rm tw5,6} = 
\left( \frac{e^{m_B^2/M^2}}{2 m_B^2 f_B} \right)
\alpha_s \langle \bar q q \rangle \frac{C_F }{N_c} \pi m_b f_\pi
\int\limits_{m_b^2}^{\infty} \! d s \! \sum_{n=2,3,4} \! \! \rho_n (q^2, s; s_0^B, M^2)
\label{eq:corr-to-vec-B-pi-FF}
\end{equation}
with the auxiliary functions $\rho_n (q^2, s; s_0^B, M^2)$ taking the form 
\begin{equation}
\rho_n (q^2, s; s_0^B, M^2) = \frac{(-1)^{n-1}}{(n-1)!}  \, g_{n} (q^2, s)
\frac{d^{n-1}}{d s^{n-1}} \left[\theta(s_0^B - s) e^{-s/M^2} \right],
\label{eq:rho-n}
\end{equation}
where the derivatives in $s$ emerge due to the higher power of the denominator in 
(\ref{eq:OPE-disp-form}), yielding the surface terms in the LCSR at $s= s_0^B$.

\section{Numerical analysis}
\begin{table}[t]\center
\begin{tabular}{|l|c|c|}
\hline
& $q^2 = 0$  & $q^2 = 10$ GeV$^2$ \\
\hline
$f_{B\pi}^+ (q^2)$ & 0.301 & 0.562 \\
\hline
Tw2 LO    & $47.5 \%$ & $48.2 \% $ \\
Tw2 NLO   & $6.9 \%$  & $5.9  \% $ \\
Tw3 LO    & $50.0 \%$ & $54.2 \% $ \\
Tw3 NLO   & $-4.6 \%$ & $-7.5 \% $ \\
Tw4 LO    & $0.2 \%$  & $-0.8 \% $ \\
\hline
Tw5 LO, fact & $-0.034 \% $ & $-0.042 \% $ \\
Tw6 LO, fact & $-0.004 \% $ & $-0.011 \% $ \\
\hline
\end{tabular}
\caption{The value of the $B \to \pi$ form factor at two typical values
$q^2 = 0$ and $q^2 = 10$~GeV$^2$ and the partial contributions 
to the LCSR.}
\label{tab:part-contr}
\end{table}

\begin{figure}[t]
\begin{center}
\includegraphics[scale=0.8]{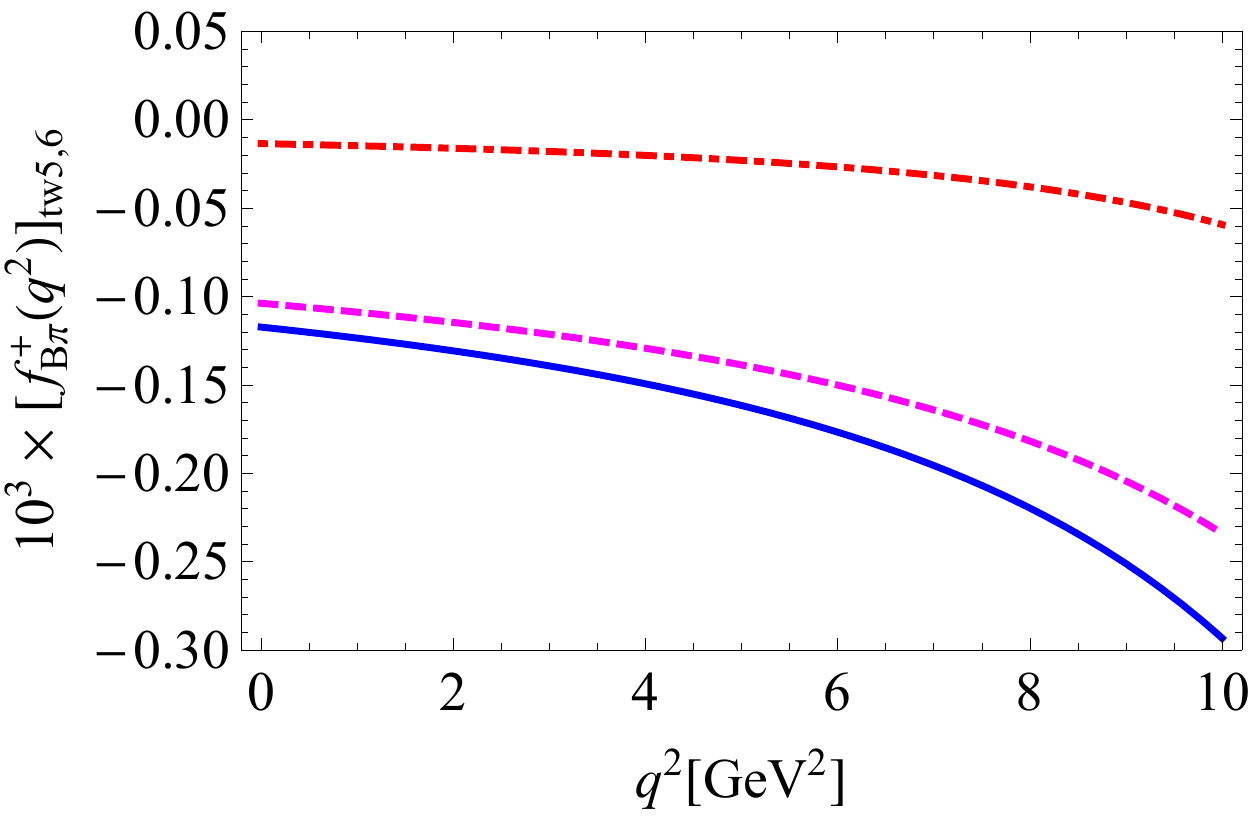}
\caption{The factorizable twist-5 and twist-6 corrections to
the vector $B\to \pi$ form factor. The dot-dashed (red) curve is the twist~6. The 
dashed (magenta) one is the twist~5 and the solid (blue) curve is the sum of the two.}
\label{fig:FFs-tw5-6-plots}
\end{center}
\end{figure}

In order to estimate the numerical impact of the factorizable twist-5 and twist-6 terms
on the vector $B \to \pi$ form factor we need to specify the input used in the LCSR. 
First of all, the values of the $B$-mesons mass $m_{B^0} = 5.27931 \, {\rm GeV}$ and 
the pion decay constant $f_\pi =130.4 \, {\rm MeV}$ are taken from \cite{PDG}.
The mass of $b$-quark is used in $\overline{MS}$-scheme and 
we adopt the interval $\overline{m}_b(\overline{m}_b)=4.18\pm 0.03$ GeV \cite{PDG}. 
The value of the quark condensate density 
$\langle \bar q q \rangle (2 \, {\rm GeV}) = - (277^{+12}_{-10} \, {\rm MeV})^3 $ 
is taken from \cite{Khodjamirian:2013}.
The normalization parameter of the twist-3 DAs $\mu_\pi$ is determined by means of
ChPT relations and we use $\mu_\pi (2 \, {\rm GeV}) = 2.50 \, {\rm GeV}$ 
following \cite{Khodjamirian:2009}.
For the renormalization scale we use the value $\mu = 3 \, {\rm GeV}$.
The $B$-meson decay constant can be extracted from the QCD sum rules
and we apply the value $f_B = 202 \, {\rm MeV}$ corresponding to the NLO accuracy
of the corresponding sum rules \cite{Khodjamirian:2013}.
Furthermore, the Borel parameter $M^2$ and the continuum threshold $s_0^B$
are taken at their typical values $M^2 = 16 \, {\rm GeV}^2$ and $s_0^B = 37.2 \, {\rm GeV}^2$
used as central values in the most recent paper \cite{Khodjamirian:2017fxg}. 

Concerning the choice of the twist-2 and twist-3 pion DAs, 
we restrict ourselves by the asymptotic form 
$\varphi (v) = 6 v \bar v$, $\varphi_p (v) = 1$ and $\varphi_\sigma (v) = 6 v \bar v$,
sufficient for our accuracy having in mind that the nonasymptotic corrections
to these DA's are relatively small.
Implementing the explicit forms for the DA's allows to perform an integration 
over $u$ in (\ref{eq:g2}) - (\ref{eq:g4}) and to determine the auxiliary 
functions $g_n (q^2, s)$ entering the LCSR for the vector $B \to \pi$ form factor
(\ref{eq:corr-to-vec-B-pi-FF}).

The numerical results for $f_{B\pi}^+ (q^2)$ corresponding to the above described input 
are presented in Fig.~\ref{fig:FFs-tw5-6-plots}, where the $q^2$-dependence
of the factorizable twist-5 and twist-6 corrections is plotted.  
Note that the corrections grow al large $q^2$ as it should be,
reflecting the growth of the higher twists effects in  the region
of low recoil, where OPE starts to diverge. 
In Tab.~\ref{tab:part-contr} we present separate contributions to the LCSR 
for the vector $B \to \pi$ form factor at two typical values 
$q^2 = 0$ and $q^2 = 10$ GeV$^2$ in order to demonstrate the magnitude of 
the factorizable higher twist corrections to the vector $B \to \pi$ form factor.
We found that in the whole domain of $q^2$ of the LCSR applicability 
the relative contributions of the higher twist effects do not exceed
$0.05 \%$ revealing their strong suppression.
The obtained result justifies a standard truncation of the OPE in LCSR up to the twist-4 terms. 
It is important to note that one of the sources of such suppression is 
a largeness of the $b$-quark mass. 
We also extended the analysis for the LCSRs for other, $B \to K$ and $B_s \to K$ 
transition vector form factors. We found that in all these cases, 
the factorizable higher twist effects are also significantly suppressed.
The corresponding corrections could have more sizeable effects in the case of 
$D \to \pi$ and $D \to K$ from factor due to a smaller value of the $c$-quark mass.
We plan to perform such analysis in the future.

\section{Conclusion}
In this paper we estimate the higher twist effects in the LCSR for the $B \to \pi$ 
vector form factor in the framework of the factorization approximation.
To this end, the light-cone expansion of the massive quark propagator including 
the higher derivatives of the gluon-field strength is derived.
The corresponding expression is in agreement with the leading order expansion
of the massive propagator \cite{Belyaev:1994zk} 
and in the massless quark limit reproduces the propagator obtained in \cite{Balitsky:1987bk}.
Our result has a more general relevance since it can be used in any other
application of LCSR where one needs the LC-expansion of the massive quark propagator.
We derive the analytical expressions for the factorizable twist-5 and twist-6
contributions to the LCSR for the vector $B \to \pi$ form factor.
The relevant numerical analysis reveals that these effects are extremely suppressed.
This justifies the conventional truncation of the operator product expansion 
in the light-cone sum rules up to twist-4 terms adopted in the previous LCSR analyses.

\section*{Acknowledgements}
I am grateful to Alexander Khodjamirian for encouraging to carry out this project 
and for the helpful discussions and careful reading of the manuscript. 
I appreciate the helpful discussion with Vladimir Braun.
The work is supported by the Nikolai-Uraltsev Fellowship of Siegen University and 
by the DFG Research Unit FOR 1873  "Quark Flavour Physics and Effective Theories",
contract No KH 205/2-2,  and partially by the Russian Foundation for Basic Research 
(project No. 15-02-06033-a).

\section*{Appendix}

In order to present the OPE result for the correlation function in the form
of dispersion integral we need to perform some transformations.
The integrals
\begin{equation}
I_n =
\int\limits_0^1 \! \int\limits_0^1 d v \,d u \, \frac{f_n (q^2, u ,v)}
{[m_b^2 - (q + u v p)^2]^n}, \quad n = 2,3,4,
\end{equation}
result from the diagram (a) of Fig.~\ref{fig:diag-tw-5-6}, and 
\begin{equation}
J_n = \int\limits_0^1 \! \int\limits_0^1 d v \,d u \,
\frac{\bar f_n (q^2, u ,v)}
{[m_b^2 - (q + (u + v - u v)p)^2]^n}, \quad n = 2,3,4, 
\end{equation}
from diagram (b).
The functions $f_n (q^2, u ,v)$ and $\bar f_n (q^2, u ,v)$ 
can be easily read off eq.~(\ref{eq:F-tw-5-6}).
Our task is to present both $I_n$ and $J_n$ in the form of dispersion integral.
To this end, in the integrals $I_n$ of the first type we replace the variable $v$ by
$\alpha = u v $, and change then the integration order 
$$
\int\limits_0^1 du \int\limits_0^u d \alpha \, (\ldots) =  
\int\limits_0^1 d \alpha \int\limits_\alpha^1 d u \, (\ldots).
$$ 
Afterwards, we introduce a new variable $s$ as follows
\begin{equation}
\alpha = \frac{m_b^2 - q^2}{s - q^2} \equiv u_1 (s, q^2).
\label{eq:u1}
\end{equation}
Finally, the integrals $I_n$ transform to 
\begin{equation}
I_n = \int\limits_{m_b^2}^\infty d s \! \! \! 
\int\limits_{u_1 (s, q^2)}^1 \! \! \! \frac{d u}{u} 
\frac{(s-q^2)^{n-2}}{(m_b^2-q^2)^{n-1}}
\frac{f_n \left(q^2, u, \displaystyle\frac{m_b^2-q^2}{u(s-q^2)}\right)}{(s-(p+q)^2)^n}.
\label{eq:int-trans-k1}
\end{equation} 
For the integrals of the second type $J_n$ we perform the replacements 
$u \to 1 - u = \bar u$ and $v \to 1 - v = \bar v$.
The next steps are similar to the previous case and the integrals $J_n$ finally 
transform to 
\begin{equation}
J_n = 
\int\limits_{m_b^2}^\infty d s \! \! \!
\int\limits_{u_2 (s, q^2)}^1 \!\! \! \frac{d \bar u}{\bar u} 
\frac{(s-q^2)^{n-2}}{(m_b^2-q^2)^{n-1}}
\frac{\bar f_n \left(q^2, \bar u, \displaystyle\frac{s - m_b^2}{\bar u (s-q^2)}\right)}
{(s-(p+q)^2)^n},
\label{eq:int-trans-k2}
\end{equation}
with $u_2 (s, q^2)$ defined as:
\begin{equation}
u_2 (s, q^2) = \frac{s - m_b^2}{s - q^2}.
\label{eq:u2}
\end{equation}
Note, that in the above integrals the dependence on the variable $(p+q)^2$ is reduced
to the denominator in the form $(s- (p+q)^2)^n$.
This significantly simplifies the derivation of the LCSR for the correlation function. 
With help of (\ref{eq:int-trans-k1}) and (\ref{eq:int-trans-k2}), 
the OPE result for the correlation function transforms to the quasi-dispersion form
(\ref{eq:OPE-disp-form}) with the functions $g_n (q^2,s)$ listed below:
\begin{eqnarray}
g_2 (q^2,s) & = & \frac{1}{m_b^2 - q^2}\int\limits_{u_1}^1 \frac{d u}{u} 
\varphi (u_1/u)
+ \frac{2}{m_b^2 - q^2}\int\limits_{u_2}^1 \frac{d u}{u} 
(1 - 2 u \bar u) \varphi (u_2/u),
\label{eq:g2} \\ 
g_3 (q^2,s) & = & 
\frac{4 q^2 (s - q^2)}{\left(m_b^2-q^2\right)^2} 
\int\limits_{u_1}^1 d u \, \bar u \, \varphi (u_1/u)
- \frac{4 \mu_{\pi} m_b}{m_b^2 - q^2} \int\limits_{u_1}^1 d u \, \bar u \, \varphi_p (u_1/u) 
\nonumber \\
& + & \frac{4 m_b^2 (s - q^2)}{\left(m_b^2-q^2\right)^2} 
\int\limits_{u_2}^1 d u \, \bar u \, \varphi (u_2/u)
+ \frac{4 \mu_{\pi} m_b}{m_b^2 - q^2} \int\limits_{u_2}^1 d u \, \bar u \, \varphi_p (u_2/u), 
\label{eq:g3} \\
g_4 (q^2,s) & = & 
2 \mu_{\pi } m_b \frac{\left(s - q^2\right)^2 \left(m_b^2+q^2\right)}{\left(m_b^2-q^2\right)^3}
\int\limits_{u_1}^1 d u u \bar u \, \varphi _{\sigma} (u_1/u) 
\nonumber \\
& + & 2 \mu_{\pi } m_b \frac{\left(s - q^2 \right)^2 \left(m_b^2+q^2\right)}{\left(m_b^2-q^2\right)^3}
\int\limits_{u_2}^1 d u u \bar u \, \varphi _{\sigma} (u_2/u),
\label{eq:g4}
\end{eqnarray}
where $u_{1,2} = u_{1,2} (s, q^2)$ are already defined in (\ref{eq:u1}) and (\ref{eq:u2}).
Inserting the explicit expressions for the pion LCDA's $\varphi (v), \varphi_p (v)$
and $\varphi_\sigma (v)$ allows to perform an integration over variable $u$ in 
(\ref{eq:g2}), (\ref{eq:g3}) and (\ref{eq:g4}).


\begin{thebibliography}{10}

\bibitem{Balitsky:1986st}
  I.I.~Balitsky, V.M.~Braun and A.V.~Kolesnichenko,
  Sov.\ J.\ Nucl.\ Phys.\  {\bf 44} (1986) 1028
  [Yad.\ Fiz.\  {\bf 44} (1986) 1582].

\bibitem{Chernyak:1990ag}
  V.L.~Chernyak and I.R.~Zhitnitsky,
  Nucl.\ Phys.\ B {\bf 345} (1990) 137.

\bibitem{Belyaev:1993wp}
  V.M.~Belyaev, A.~Khodjamirian and R.~Ruckl,
  Z.\ Phys.\ C {\bf 60} (1993) 349, hep-ph/9305348.

\bibitem{Belyaev:1994zk}
  V.M.~Belyaev, V.M.~Braun, A.~Khodjamirian and R.~Ruckl,
  Phys.\ Rev.\ D {\bf 51} (1995) 6177, hep-ph/9410280.

\bibitem{Duplancic:2008ix}
  G.~Duplancic, A.~Khodjamirian, T.~Mannel, B.~Melic and N.~Offen,
  JHEP {\bf 0804} (2008) 014,
  arXiv:0801.1796 [hep-ph].

\bibitem{Khodjamirian:1997ub}
  A.~Khodjamirian, R.~Ruckl, S.~Weinzierl and O.I.~Yakovlev,
  Phys.\ Lett.\ B {\bf 410} (1997) 275,
  hep-ph/9706303.
  
\bibitem{Bagan:1997bp}
  E.~Bagan, P.~Ball and V.M.~Braun,
  Phys.\ Lett.\ B {\bf 417} (1998) 154,
  hep-ph/9709243.
  
\bibitem{Ball:2001fp}
  P.~Ball and R.~Zwicky,
  JHEP {\bf 0110} (2001) 019,
  hep-ph/0110115.

\bibitem{Bharucha:2012wy}
  A.~Bharucha,
  JHEP {\bf 1205} (2012) 092,
  arXiv:1203.1359 [hep-ph].

\bibitem{Balitsky:1987bk}
  I.I.~Balitsky and V.M.~Braun,
  Nucl.\ Phys.\ B {\bf 311} (1989) 541.
  
\bibitem{Braun:1999uj}
  V.M.~Braun, A.~Khodjamirian and M.~Maul,
  Phys.\ Rev.\ D {\bf 61} (2000) 073004,
  hep-ph/9907495.
  
\bibitem{Agaev:2010aq}
  S.S.~Agaev, V.M.~Braun, N.~Offen and F.A.~Porkert,
  Phys.\ Rev.\ D {\bf 83} (2011) 054020,
  arXiv:1012.4671 [hep-ph].

\bibitem{Colangelo:2000dp}
  P.~Colangelo and A.~Khodjamirian,
  In *Shifman, M. (ed.): At the frontier of particle physics, vol. 3* 1495-1576,
  hep-ph/0010175.

\bibitem{PDG}
   C.~Patrignani {\it et al.} [Particle Data Group],
   Chin.\ Phys.\ C {\bf 40} (2016), 100001.

\bibitem{Khodjamirian:2013}
  P.~Gelhausen, A.~Khodjamirian, A.A.~Pivovarov and D.~Rosenthal,
   Phys.\ Rev.\ D {\bf 88} (2013) 014015,
   Erratum: [Phys.\ Rev.\ D {\bf 89} (2014) 099901],
   Erratum: [Phys.\ Rev.\ D {\bf 91} (2015) 099901],
  arXiv:1305.5432 [hep-ph].

\bibitem{Khodjamirian:2009}
  A.~Khodjamirian, C.~Klein, T.~Mannel and N.~Offen,
  Phys.\ Rev.\ D {\bf 80} (2009) 114005,
  arXiv:0907.2842 [hep-ph].

\bibitem{Khodjamirian:2017fxg}
  A.~Khodjamirian and A.V.~Rusov,
  arXiv:1703.04765 [hep-ph].

\end{thebibliography}
\end{document}